\documentclass[%
 reprint,
 superscriptaddress,
 amsmath,amssymb,
 aps,
 prl,
]{revtex4-2}

\usepackage{graphicx}% Include figure files
\usepackage{dcolumn}% Align table columns on decimal point
\usepackage{bm}% bold math
\usepackage{textcomp}
\usepackage{mathptmx}
\usepackage[colorlinks=true,allcolors=blue,urlcolor=black]{hyperref}

\usepackage{xr}
\makeatletter
\newcommand*{\addFileDependency}[1]{
	\typeout{(#1)}
	\@addtofilelist{#1}
	\IfFileExists{#1}{}{\typeout{No file #1.}}
}
\makeatother

\newcommand*{\myexternaldocument}[1]{
	\externaldocument{#1}
	\addFileDependency{#1.tex}
	\addFileDependency{#1.aux}
}
\myexternaldocument{supplement}
\begin{document}

%\preprint{APS/123-QED}

\title{Fractal and knot-free chromosomes facilitate nucleoplasmic transport}

\author{Yeonghoon Kim}
\affiliation{Department of Physics, Pohang University of Science and Technology, Pohang 37673, Republic of Korea}
\author{Ludvig Lizana}%
\email[]{ludvig.lizana@umu.se}
\affiliation{Integrated Science Lab, Department of Physics, Ume\aa~University, Ume\aa~ 90187, Sweden
}
\author{Jae-Hyung Jeon}
\email[]{jeonjh@postech.ac.kr}
\affiliation{Department of Physics, Pohang University of Science and Technology, Pohang 37673, Republic of Korea}
\affiliation{Asia Pacific Center for Theoretical Physics, Pohang 37673, Republic of Korea}

\date{\today}

\begin{abstract}
Chromosomes in the nucleus assemble into hierarchies of 3D domains that, during interphase, share essential features with a knot-free condensed polymer known as the fractal globule (FG). The FG-like chromosome likely affects macromolecular transport, yet its characteristics remain poorly understood. Using computer simulations and scaling analysis, we show that the 3D folding and macromolecular size of the chromosomes determine their transport characteristics. Large-scale subdiffusion occurs at a critical particle size where the network of accessible volumes is critically connected. Condensed chromosomes have connectivity networks akin to simple Bernoulli bond percolation clusters, regardless of the polymer models. However, even if the network structures are similar, the tracer's walk dimension varies. It turns out that the walk dimension depends on the network topology of the accessible volume and dynamic heterogeneity of the tracer's hopping rate. We find that the FG structure has a smaller walk dimension than other random geometries, suggesting that the FG-like chromosome structure accelerates macromolecular diffusion and target-search.

\end{abstract}

\maketitle

The nucleus is the central organelle in a cell for its living processes from the viewpoint of the central dogma. Its interior is filled with highly-packed chromosomal DNAs and a plethora of poly-dispersed macromolecules with sizes ranging from a few nanometers to sub-micrometers. In processes such as gene regulation, DNA repair, and epigenetics, these macromolecules must explore the chromosome-made labyrinth space to find the specific target sites or counterpart molecules. Novel experiments using a chromosome conformation capture technique (e.g., Hi-C) have shown that eukaryotic chromosomes during interphase fold into hierarchical and fractal condensates, far from random coils, possessing territory-preserving and knot-free folding structures, as well as many loops ~\cite{lieberman-aiden,grosberg_prog}. The fractal globule (FG) was proposed as a minimal chromosome model~\cite{lieberman-aiden,mirny_review} that captures these essential features of the human chromosome~\cite{grosberg_prog,tamm2015anomalous}. It is a space-filling, knot-free, and self-similar polymer condensate~\cite{tamm2015anomalous,grosberg_FG}. 

As a theoretical model, we consider nucleoplasmic transport as diffusion in polymer condensates. To date, a typical approach was to depict the nucleoplasmic environment as a random structure in terms of continuum percolation~\cite{franosch_review,rippe}, fractal space~\cite{ellenberg,benichou2011prl}, or random polymer architectures~\cite{langowski_jcp}. These studies, albeit insightful, neglected the folding complexity in the chromosome condensates that can arguably determine the transport dynamics by reshaping the particle's accessible volume. A prominent example from an in vitro experiment is a tracer particle diffusing through a reconstituted actin network. With the variation of the particle-to-mesh size ratio, the tracers exhibited various diffusion patterns from a continuous-time random walk with a power-law trapping time statistics to a Fickian yet non-Gaussian diffusion~\cite{wong,granick_actin_fick}. The macromolecules inside a nucleus also showed that they diffuse differently depending on location (i.e., chromosome-occupying or -free region) and physical size~\cite{ellenberg, fecko, rippe, tyagi, funatsu, enquist, lichter,garini_eli,brangwynne}. 

Based on extensive simulations and scaling theory, in this Letter, we demonstrate that the chromosome's native folding structure does not merely hinder the particle diffusion but plays a crucial role in transporting the macromolecules across the nucleus. Using FGs as a proxy for the interphase human chromosome, we explicitly simulate the molecular transport therein for various tracer sizes. We find that the tracer suffers an abrupt geometrical change of the accessible volume as its size increases to a critical value. For tracers in the vicinity of the critical size, the transport shows a structure-dependent critical subdiffusion or otherwise is Fickian and structure-invariant. Surprisingly, the walk dimension from the FG is much smaller than  from the equilibrium globule and melted linear chain. This suggests that the FG's self-similarly folded structure, and thus chromosomes, generates an effective accessible space for large macromolecules to diffuse over longer distances than in other random geometries. With such enhanced transport mechanism, the macromolecules can effectively find their target sites or escape the nucleus.       

\textit{The model}.---We construct the polymers on a 3D cubic lattice (lattice constant, $a=1$) where we set the fraction of occupied edges to $\Pi=0.3$ ($\Pi=1/3$ is the space-filling limit). We consider two polymer models: fractal globule (FG) and melted linear chain (MLC) (Fig.~1(a) \& S1). The FG serves as a mammalian chromosome model (with the territory-preserving and knot-free characteristics), whereas the MLC is a well-mixed entangled polymer condensate~\cite{grosberg_prog,mirny_review}. We construct the FG using the conformation-dependent polymerization algorithm~\cite{tamm2015anomalous,SH_community}. It generates meandering paths that return close to the previously visited sites without forming knots. To create the MLC, we simulate a non-biased self-avoiding random walk in a closed volume. The MLCs share structural properties with equilibrium globules (Fig.~S3). We also study a random bond percolation clusters (non-polymeric) and Moore curves (mathematical fractal) serving as comparisons (see Supplementary Material \cite{supplement}, Figs.~S1--S3).

We represent the polymer segments as hemisphere-capped cylinders of radius $R_\mathrm{obs}$ (Fig.~S1(b)). They exclude the volume of the embedded tracer of a radius $R$. Due to the excluded volume (EV) interaction, we map the system onto a point-particle problem with an effective radius $R_\mathrm{eff}=R+R_\mathrm{obs}$, which is the key parameter in our model. Because the chromosome dynamics is much slower than the macromolecular diffusion studied here \cite{garini_eli}, we assume that the obstacles are immobile. Effects of dynamic FGs are discussed in \cite{supplement}.
To generate tracer trajectories, we integrate the overdamped Langevin equation for a given obstacle architecture and $R_\mathrm{eff}$. The EV interaction is assumed to be hard-core and treated kinematically via rejection sampling~\cite{cichocki1990dynamic}; see \cite{supplement} for the full description.

%%%%%%%%%%%%%%%%%%%%%%%% FIG 1 %%%%%%%%%%%%%%%%%%%%%%%%%
\begin{figure}[t]
    \centering
    \includegraphics{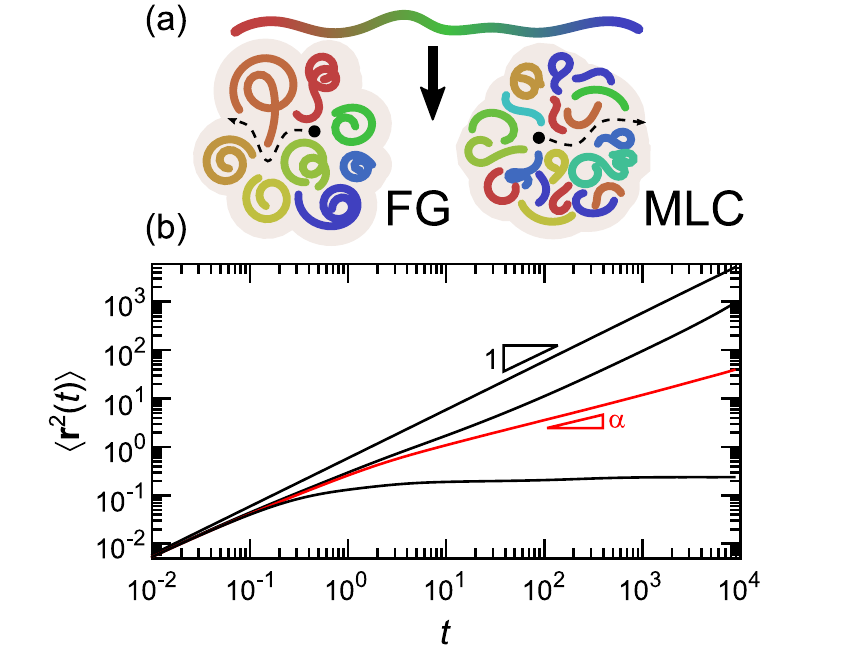}
    \caption{
    (a) Polymer configurations: fractal globule and melted linear chain. Red-to-blue color indicates increasing monomer index. Dots represent macromolecules exploring the FG and MLC geometries.
    (b) Tracer MSDs in FGs. From top to bottom, $R_\mathrm{eff}=0.100$, $0.600$, $0.630$ and $0.800$. The red curve shows the critical subdiffusion with $\alpha\simeq 0.5$.
    } 
    \label{fig1}
\end{figure}
%%%%%%%%%%%%%%%%%%%%%%%%%%%%%%%%%%%%%%%%%%%%%%%%%%%%%%

\textit{Geometry-induced anomalous diffusion}.---Nucleoplasmic transport of macromolecules was experimentally investigated using proteins \cite{ellenberg,fecko,rippe}, mRNAs \cite{tyagi,funatsu}, viral capsids \cite{enquist}, and organelles \cite{lichter}. These experiments showed that the mean-squared displacement (MSD) increases with time as $\langle \mathbf{r}^2(t)\rangle \propto t^\alpha$ where $\alpha\in (0,1]$ is the anomalous exponent~\cite{ralf_PCCP2014}. Importantly, $\alpha$ is system-specific, and depends appreciably on the macromolecular size and the chromosomes' folding status.

To understand such dynamic heterogeneity in the experiments, we focus on how the tracer diffusion changes with its size $R_\mathrm{eff}$. Figure~1(b) shows representative MSD curves in FGs for increasing $R_\mathrm{eff}$. Each curve corresponds to the time- and ensemble-averaged MSD over $10^2$ obstacle configurations and $10^2$ trajectories, each $10^7$ time steps long.

The plot shows that the tracer experiences different geometries that depends on its size. Studying sample trajectories supports this conclusion (Fig.~S4). We find three key observations: (1) For sufficiently small sizes ($R_\mathrm{eff}=0.1)$, the tracer exhibits a Fickian diffusion as in free space. (2) Tracers larger than this limit interact with the obstacle at the length- or time-scales of $\langle \mathbf{r}^2(t)\rangle\sim a^2$, giving rise to a transient subdiffusion ($R_\mathrm{eff}=0.6$). Then, the tracers recover a Fickian diffusion with a reduced diffusivity. (3) Remarkably, when $R_\mathrm{eff}$ is close to a critical value $R_\mathrm{cr}=0.625$, the tracers have a several-decade long subdiffusion regime with $\alpha \simeq 0.5$ (red line) at the length-scales of $\langle \mathbf{r}^2\rangle \gtrsim a^2$. This is reminiscent of anomalous diffusion on a critical percolation clusters~\cite{franosch_review,bunde2012fractals}. The tracer appears as hopping between the nearest cavities through a narrow hole (Fig.~S4). (4) If the tracers are larger than $R_\mathrm{cr}$, they cannot access the whole space but undergo confined diffusion ($R_\mathrm{eff}=0.8$).   The tracer transport in the MLC is similar to that in the FG, but with a smaller exponent $\alpha$ at $R_\mathrm{eff}\simeq R_\mathrm{cr}$ (Fig.~S5).

To quantify the relationship between the anomalous exponent, tracer size, and obstacle geometry, we calculate how the MSD slope changes over time, defining the smallest slope as $\alpha=\alpha_\mathrm{min}$ (Fig.~S6). In Fig.~2(a), we show how $\alpha$ changes with $R_\mathrm{eff}$ for the FG and MLC. The plot shows that: 
(1) $\alpha$ tends to decrease as $R_\mathrm{eff}$ increases.
(2) $\alpha$ decays in the same manner regardless of the obstacle geometry for the tracers of $R_\mathrm{eff}< R_\mathrm{cr}$.
(3) $\alpha$ drops significantly for the tracers of $R_\mathrm{eff}\gtrsim R_\mathrm{cr}$, where the diffusion dynamics differ depending on the obstacle geometry.

%%%%%%%%%%% FIG 2 %%%%%%%%%%%%%%%
\begin{figure*}[th!]
    \centering
    \includegraphics{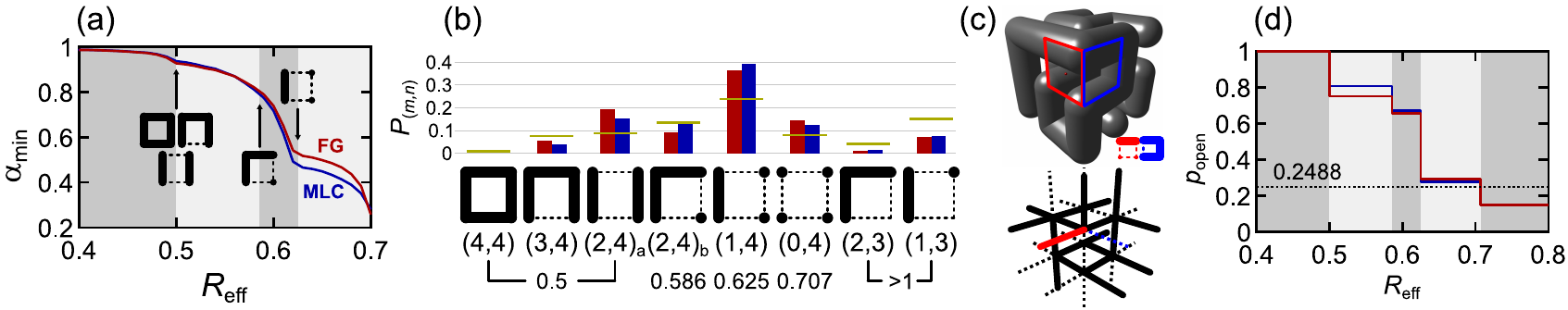}
    \caption{
    (a) Anomalous exponent ($\alpha=\alpha_\mathrm{min}$) vs. $R_\mathrm{eff}$. The borders between shaded regions show the threshold $R_\mathrm{eff}$s when some EV motifs close.
    (b) The representative eight EV motifs and their occurrence probability $P_{(m,n)}$, where $m$ and $n$ are the numbers of blocked edges and vertices. Below each motif, we denote the threshold $R_\mathrm{eff}$ when the motif closes. The complete list of 14 EV motifs is depicted in Fig.~S7.
    (c) Example showing how we create the accessible lattice (AL): (Top) an obstacle geometry with two exemplified motifs, $(1,4)$ and $(3,4)$. If $0.5<R_\mathrm{eff}<0.625$, the $(1,4)$ is open (red) but the $(3,4)$ (blue) closed. (Bottom) The AL in red marks the open $(1,4)$ motif.  
    (d) Fraction of open motifs $p_\mathrm{open}$ as a function $R_\mathrm{eff}$. The dotted line shows $p_c$ for the Bernoulli bond percolation (cubic lattice).
    } 
    \label{fig2}
\end{figure*}
%%%%%%%%%%%%%%%%%%%%%%%%%%

\textit{EV motifs.}---To gain a physical understanding of these findings, we study the local obstacle configurations that restrict the tracer while moving between cubic cells. 
We find 14 possible variants that we call excluded volume (EV) motifs (cartoons in Fig.~2(b) \& S7). We label these motifs with indices $(m,n)$ indicating the number of obstacle-occupied edges $(m)$ and vertices $(n)$. Figure~2(b) also shows the occurrence probability $P_{(m,n)}$ for each EV motif; red and blue bars correspond to the FG and MLC, respectively. The horizontal lines are $P_{(m,n)}$ for random bond percolation serving as a reference case we derived theoretically (see the expressions in \cite{supplement}). 

Apart from overall systematic differences between the FG and MLC, we see that the $(1,4)$-motif is the most abundant for both geometries. Additionally, below the icon for each motif, we indicate the $R_\mathrm{eff}$ values when they get closed.  For example, if $R_\mathrm{eff}\gtrsim 0.5$, three motifs--$(4,4)$,  $(3,4)$, and $(2,4)_a$--become sealed off. This causes the tracer to make detours because every instance of $(4,4)$,  $(3,4)$, or $(2,4)_a$ that belonged to a path is cut. If the tracer becomes even larger, the diffusive dynamics will not significantly change until $R_\mathrm{eff}=0.586$ is reached. At this point, the $(2,4)_b$ motif also closes. This analysis explains why $\alpha$ decays as $R_\mathrm{eff}$ grows and why this occurs in steps. Importantly, we find that the critical subdiffusion occurs (in Fig.~2(a) with $\alpha\simeq 0.5$) when the $(1,4)$-motif gets closed at $R_\mathrm{eff}=0.625$. We observe the same behavior in MLCs. Thus, for  chromosome-like obstacles, $R_\mathrm{cr}$ has a geometrical interpretation of the most abundant motif being closed. 

\textit{The accessible lattice and  its percolation structure}.---To better understand the relationship between the obstacle geometry and the transport dynamics, we construct the accessible lattice (AL). An AL is a bond percolation cluster where the vertex symbolizes the unit cell's center. Two vertices are connected if the EV motif separating them is open (see Fig.~2(c) for an example). The AL represents the network of possible tracer paths that depends on the tracer's size and obstacle geometry.

The key parameter that characterizes the AL is the percolation density $p_\mathrm{open}$ or the fraction of open EV motifs. Similar to the trend for $\alpha$ in Fig.~2(a), Fig.~2(d) shows that $p_\mathrm{open}$ has a staircase decrease with increasing $R_\mathrm{eff}$~\cite{note1}. As noted above, $\alpha$ decreases significantly when the (1,4)-motif becomes blocked at $R_\mathrm{eff}\approx R_\mathrm{cr}$. This occurs close to the critical bond density $p_c\approx 0.2488$ associated with the random Bernoulli percolation cluster. The critical connectivity expects the transport should be subdiffusive, as justified by the observed onset of the critical subdiffusion near $R_\mathrm{eff}=0.625$.

Notably, when $p_\mathrm{open}\approx p_c$, the FG and MLC have very similar ALs, which are akin to the random Bernoulli percolation. The largest AL cluster in both FGs and MLCs, has the same fractal dimension as in Bernoulli percolation $d_f\approx 2.5$ (Fig.~S8(a)); the ALs have the same cluster size distribution (FG, Fig.~S8(b) and MLC, Fig.~S8(c)) as the Bernoulli percolation at $p=p_\mathrm{open}$; additionally, the scaling between Euclidean and chemical distance (Fig.~S8(d)) and the two-point correlation (Fig.~S8(e)) are similar to each other. For a complete analysis, see Figs.~S8~\&~S9 and accompanying text~\cite{supplement}. 

\textit{Walk dimensions and their geometry dependence}.---The AL analysis suggests that the tracers have the same critical subdiffusion in FGs and MLCs with the walk dimension (or $\alpha$) predicted by the random walk on a Bernoulli percolation. However, the walk dimension is non-universal and depends on the obstacle organization. To demonstrate this, we use results from random percolation theory, where the MSD for a random walker obeys the scaling relation~\cite{bunde2012fractals}
\begin{equation}\label{eqn:msdscale}
\langle {\mathbf{r}^2(t)} \rangle = t^{2/d_w'} F(t/t_\xi)
\end{equation}
where $d_w'$ is the walk dimension for tracers from both infinite and finite percolation clusters, and $F(x)$ is a scaling function. $d_w'$ is related to the anomalous exponent by $\alpha = 2/d_w'$, and  $F(x)\propto 1$ for $x\ll1$ and $F(x) \propto x^{1-2/d_w'}$ for $x\gg1$. For $t>t_\xi$, we expect to see Fickian diffusion. 

To extract $d_w'$, we use Eq.~\eqref{eqn:msdscale} to rescale the simulated MSD curves and vary $d_w'$ to maximize the data collapse. In Fig.~3(a), we show the result for the FG where $R_\mathrm{eff}$s are close to $R_\mathrm{cr}$. For large times, the collapse is excellent when $d_w'=3.84(\pm 0.23)$. However, there is some disagreement at short times ($t\ll t_\xi$). This regime corresponds to diffusion over the length-scales shorter than the lattice constant where we do not expect Eq.~\eqref{eqn:msdscale} to hold.

The same scaling form holds for the MLC, but the exponent deviates significantly from the FG's, $d_w'=5.71(\pm0.94)$ (Fig.~S10). Using $d_w'$ we estimate the anomalous exponent at the critical condition $R_\mathrm{eff}\approx R_\mathrm{cr}$ via $\alpha=2/d_w'$, giving $\alpha = 0.52$ (FG) and $0.35$ (MLC). These values agree with the estimate of the MSD slopes that we obtained from Fig.~1(b) and Fig.~S5. We also study $d_w'$ in the random bond percolation (Fig.~S11), but we cannot collapse the data for a single value $d_w'$.

To cross-check the walk dimensions, we measure them in a different way using a scaling relation for the displacement PDF in the anomalous regime of $t<t_\xi$ (Fig.~3(b))~\cite{bunde2012fractals, havlin1985probability}:
\begin{equation}\label{eqn:pdfscaleours}
rP(r|t) \propto (r/t^{1/d_w})^{d_f} \exp[-A(r/t^{1/d_w})^{d_w/(d_w-1)}].
\end{equation}
Here, $r=|\mathbf{r}|$, $d_f(=2.52)$ is the fractal dimension, $d_w$ is the infinite-cluster walk dimension (empirically, $d_w=0.76d_w'$~\cite{bunde2012fractals}), and $A$ is a constant. Using simulated data, we calculate $P(r|t)$ and rescale it according to Eq.~\eqref{eqn:pdfscaleours} (see details in~\cite{supplement}). By varying $d_w$, we find the best collapse to be $d_w=3.49(\pm0.04)$ (FG) and $5.62(\pm 0.26$) (MLC), which is consistent what we found using Eq.~\eqref{eqn:msdscale} via $d_w\approx 0.76d_w'$. Repeating this analysis for the annealed Moore curve, we confirm that the critical diffusion we observe in FGs are independent of the preparation method (Fig.~S12 \& S13 and Sec.~IX in \cite{supplement}).
Therefore, we conclude that the transport dynamics  show a strong dependence on the chromosome structure.  

%%%%%%%%%%% FIG 3 %%%%%%%%%%%%%%%
\begin{figure}[t]
\centering
\includegraphics{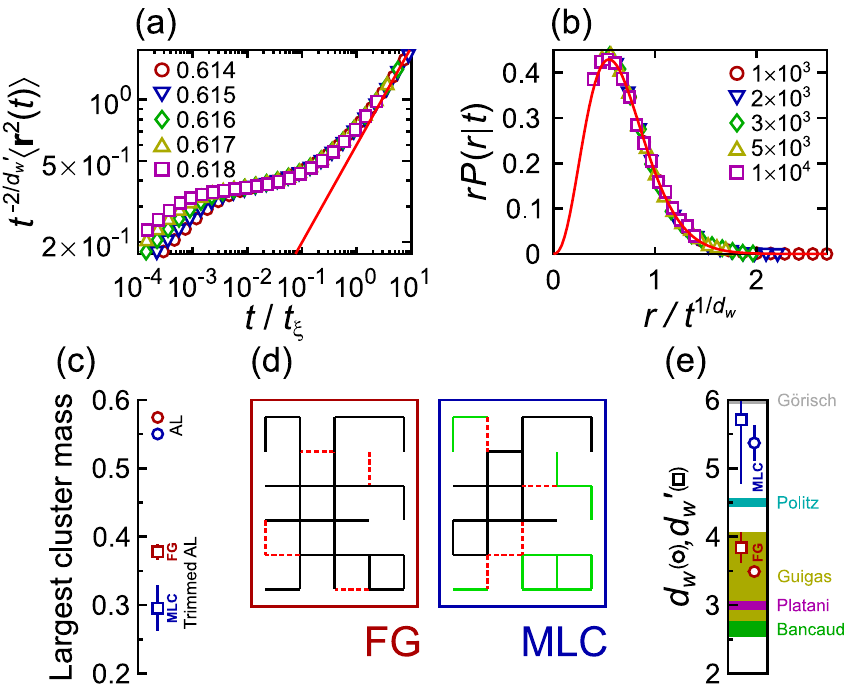}
\caption{
(a) Rescaled MSDs using Eq.~\eqref{eqn:msdscale} for five different $R_\mathrm{eff}$s in FGs. This analysis yields $d_w'=3.84$.
The solid line depicts the expected scaling at $t\to\infty$.  (b) Rescaled displacement PDFs $P(r|t)$ using Eq.~\eqref{eqn:pdfscaleours} at five lag times and $R_\mathrm{eff}=0.63$ in FGs. The best collapse yields $d_w=3.49$ along with the corresponding fit (line). 
(c) Mass fraction of the largest cluster in the FG (red) and MLC (blue) at $R_\mathrm{eff}\approx R_\mathrm{cr}$. The symbol represents the results for the original ALs ($\circ$) and the ALs after the weak edges are removed ($\square$). 
Details are described in~\cite{supplement}. 
(d) Schematics explaining the difference in the walk dimensions between the FG and MLC geometries. The black, red, and green lines constitute the original AL and correspond to the spanning cluster, weak edges, and isolated small clusters, respectively. The small clusters are the fraction of the AL connected to the spanning cluster via the weak edge. 
(e) The walk dimensions obtained from (a) and (b) are shown for the FG and MLC, with the speculated $d_w'=2/\alpha$ from several experiments \cite{ellenberg,platani,guigas2007,politz,lichter}. 
} 
\label{fig3}
\end{figure}
%%%%%%%%%%%%%%%%%%%%%%%%%%%%%%%

However, we find that the AL--the network structure of accessible volumes--is insufficient to explain that the $d_w'$s are non-universal. We calculate the MSD from simulated random walks on the ALs taken from the generated FGs and MLCs at $R_\mathrm{eff}\approx R_\mathrm{cr}$. These MSDs are found to be the same as those from random walks on the Bernoulli percolation at $p=p_\mathrm{open}$ (Fig.~S14). The AL model gives $\alpha_\mathrm{AL}\approx 0.63$ (FG) and $\approx0.58$ (MLC). This difference only reflects different $p_\mathrm{open}$. The $\alpha_\mathrm{AL}$s are not consistent with the $\alpha$s (or $2/d_w'$) in the FG and MLC.

One reason that the diffusion dynamics is not fully explained by the AL is because it does not incorporate the dynamic heterogeneity in the tracer's hopping rates. Close to the critical point $R_\mathrm{eff}\approx R_\mathrm{cr}$, we notice that there is a broad sojourn-time distribution when tracers move into connected AL sites (Fig.~S15). This means that edges have a spectrum of hopping rates that are associated with the EV motifs' open area $S_{(m,n)}$. Based on $S_{(m,n)}$, we define edge strengths (see \cite{supplement}) and find that there are several weak edges ($0<S_{(m,n)}\ll a^2$) that hardly allow tracers to pass.  

In Fig.~3(c), we compare the mass fraction of the largest (spanning) cluster in the FG and MLC before and after trimming the weak edges (see Fig.~S16 as an example). Before trimming ($\circ$), the ALs of the FG and MLC have spanning clusters of comparable mass and a similar number of weak edges. However, after removing the weak edges ($\square$), the FG's spanning cluster becomes significantly larger than the MLC's. This suggests the mechanism of the better transport efficiency of the FG-like chromosomes compared to MLCs, with a smaller $d_w'$. Fig.~3(d) recapitulates this idea. Here, we schematically depict the spanning cluster (black), the weak edges (red), and small clusters (green) connected to the spanning cluster in the AL. In the MLC, the weak edges topologically act as a bottleneck between the spanning and the small clusters. By contrary, the FG's hierarchical folding renders the accessible volume to contain less bottlenecks. This leads to a larger spanning cluster and thus enhances diffusion. 

We note that the critical subdiffusion and the enhancement of diffusion in FG geometries are preserved even if our FG geometries are dynamic or irregularly patterned in terms of EV motifs. See our further simulation studies (Fig.~S17 \& S18, Sec.~X \& XI) in \cite{supplement}.  We emphasize that the physical mechanisms explaining our findings differ from \cite{JanGrosberg2015} who studied point-particles exploring crumpled globules with weak absorption.

\textit{Implications for biological systems.}---In Fig.~3(e), we compare the walk dimensions from the FG and MLC to five experimental values (Tab.~S1 shows thirteen~\cite{supplement}). These experiments tracked diffusing macromolecules of different sizes in cell nuclei and measured $d_w'(=2/\alpha)$. While most values appear to lie between 2 and 4, they cover a broad spectrum--at least $\sim 2$--$6$. Such diversities may result from multiple mechanisms such as chromatin dynamics \cite{garini_eli,polovnikov2018} and their structural heterogeneity~\cite{ellenberg}. Nevertheless, we find a few results that are close to $d_w'$ for the FG \cite{guigas2007} and MLC~\cite{lichter}.

We relate our results to the sizes of real macromolecules and speculate regarding their diffusion dynamics. We assume that our FG model represents the 11-nm beads-on-a-string chromatin, and the lattice constant is roughly the chromatin's Kuhn length which ranges $15$--$60$~nm~\cite{hajjoul2013,cui2000}. Under such conditions, we expect that  average-sized proteins (of radius $2$~nm) have $R_\mathrm{eff}\approx 0.13$--$0.5$. For molecules of radii $<1$~nm (e.g., ATPs and nucleotides), their diffusion are almost Fickian, and are hardly affected by the chromosome structure. We expect to see critical subdiffusive behavior when the macromolecule radii are between $5$~nm (e.g., RNA polymerase) and $25$~nm (e.g., RNAs and small protein aggregates). This estimate suggests that the molecular diffusion inside the nucleus is size-sensitive and that nano- to sub-micron-sized bio-molecules exhibit various diffusion dynamics.  Indeed, this conclusion in in agreement with an experiment \cite{guigas2007} showing that 5-nm gold particles immersed in nucleoplasms exhibited subdiffusion with $\alpha\approx 0.5$--$0.6$.

To conclude, we studied particle diffusion in chromosome-like polymer structures to better understand how a DNA-filled nucleus may cause large proteins to subdiffuse. As such, any deviation from our theory indicates the presence of additional mechanisms, such as interactions with other macromolecules, binding to the polymer's segments, or non-spherical tracer shapes. Therefore, our work is useful to researchers studying protein-DNA target search, protein aggregation, or diffusion-limited gene regulatory circuits.

\begin{acknowledgments}
This work was supported by the National Research Foundation (NRF) of Korea (No.~2017K1A1A2013241) and the Swedish Foundation for International Cooperation in Research and Higher Education (STINT) (No.~KO2015-6452). We thank Eli Barkai and Jaeoh Shin for critical feedback on the manuscript. 
\end{acknowledgments}


\begin{thebibliography}{99}
\bibitem{lieberman-aiden} E. Lieberman-Aiden \textit{et al.}, Science \textbf{326}, 289 (2009).
\bibitem{grosberg_prog} J. D. Halverson, J. Smrek, K. Kremer, and A. Y. Grosberg, Rep. Prog. Phys. \textbf{77}, 022601 (2014).
\bibitem{mirny_review} L. A. Mirny, Chromosome Res. \textbf{19}, 37 (2011).
\bibitem{tamm2015anomalous} M. V. Tamm, L. I. Nazarov, A. A. Gavrilov, and A. V. Chertovich, Phys. Rev. Lett. \textbf{114}, 178102 (2015).
\bibitem{grosberg_FG} A. Y. Grosberg, S. K. Nechaev, and E. I. Shakhnovich, J. Phys. \textbf{49}, 2095 (1988).
\bibitem{franosch_review} F. H{\"o}fling and T. Franosch, Rep. Prog. Phys. \textbf{76}, 046602 (2013).
\bibitem{rippe} M. Baum, F. Erdel, M. Wachsmuth, and K. Rippe, Nat. Commun. \textbf{5}, 1 (2014).
\bibitem{ellenberg} A. Bancaud, S. Huet, N. Daigle, J. Mozziconacci, J. Beaudouin, and J. Ellenberg, EMBO J. \textbf{28}, 3785 (2009).
\bibitem{benichou2011prl} O. B{\'e}nichou, C. Chevalier, B. Meyer, and R. Voituriez, Phys. Rev. Lett. \textbf{106}, 038102 (2011).
\bibitem{langowski_jcp} C. C. Fritsch and J. Langowski, J. Chem. Phys. \textbf{133}, 07B602 (2010).
\bibitem{wong} I. Y. Wong, M. L. Gardel, D. R. Reichman, Eric R. Weeks, M. T. Valentine, A. R. Bausch, and D. A. Weitz, Phys. Rev. Lett. \textbf{92}, 178101 (2004).
\bibitem{granick_actin_fick} B. Wang, S. M. Anthony, S. C. Bae, and S. Granick, Proc. Natl. Acad. Sci. U.S.A. \textbf{106}, 15160 (2009).
\bibitem{fecko} M. K. Daddysman and C. J. Fecko, J. Phys. Chem. B \textbf{117}, 1241 (2013).
\bibitem{tyagi} D. Y. Vargas, A. Raj, S. A. Marras, F. R. Kramer, and S. Tyagi, Proc. Natl. Acad. Sci. U.S.A. \textbf{102}, 17008 (2005).
\bibitem{funatsu} Y. Ishihama and T. Funatsu, Biochem. Biophys. Res. Commun. \textbf{381}, 33 (2009).
\bibitem{enquist} J. B. Bosse, I. B. Hogue, M. Feric, S. Y. Thiberge, B. Sodeik, C. P. Brangwynne, and L. W. Enquist, Proc. Natl. Acad. Sci. U.S.A. \textbf{112}, E5725 (2015).
\bibitem{lichter} S. M. G{\"o}risch, M. Wachsmuth, C. Ittrich, C. P. Bacher, K. Rippe, and P. Lichter, Proc. Natl. Acad. Sci. U.S.A. \textbf{101}, 13221 (2004).
\bibitem{garini_eli} I. Bronstein, Y. Israel, E. Kepten, S. Mai, Y. Shav-Tal, E. Barkai, and Y. Garini, Phys. Rev. Lett. \textbf{103}, 018102 (2009).
\bibitem{brangwynne} D. S. W. Lee, N. S. Wingreen, and C. P. Brangwynne, Nat. Phys. (2021).
\bibitem{SH_community} S. H. Lee, Y. Kim, S. Lee, X. Durang, P. Stenberg, J.-H. Jeon, and L. Lizana, Sci. Rep. \textbf{9}, 6859 (2019).
\bibitem{supplement} See Supplemental Material at [url].
\bibitem{cichocki1990dynamic} B. Cichocki and K. Hinsen, Physica A \textbf{166}, 473 (1990).
\bibitem{ralf_PCCP2014} R. Metzler, J.-H. Jeon, A. G. Cherstvy, and E. Barkai, Phys. Chem. Chem. Phys. \textbf{16}, 24128 (2014).
\bibitem{bunde2012fractals} A. Bunde and S. Havlin, \emph{Fractals and disordered systems} (Springer Science \& Business Media, 2012).
\bibitem{note1} Our supplementary study shows that the decrease in $p_\mathrm{open}$ becomes smooth while the critical subdiffusion is preserved if the on-lattice obstacles are randomized in position to have irregular EV motifs. For further information, see Sec.~XI and Fig.~S18 in the Supplementary Material.
\bibitem{havlin1985probability} S. Havlin, D. Movshovitz, B. Trus, and G. H. Weiss, J. Phys. A Math. Theor. \textbf{18}, L719 (1985).
\bibitem{JanGrosberg2015} J. Smrek and A. Y. Grosberg, Phys. Rev. E \textbf{92}, 012702 (2015).
\bibitem{polovnikov2018} K. Polovnikov, M. Gherardi, M. Cosentino-Lagomarsino, and M. Tamm, Phys. Rev. Lett. \textbf{120}, 088101 (2018).
\bibitem{guigas2007}  G. Guigas, C. Kalla, and M. Weiss, Biophys. J. \textbf{93}, 316 (2007).
\bibitem{politz} J. C. R. Politz, R. A. Tuft, and T. Pederson, Mol. Biol. Cell \textbf{14}, 4805 (2003).
\bibitem{platani} M. Platani, I. Goldberg, A. I. Lamond, and J. R. Swedlow,
Nat. Cell Biol. \textbf{4}, 502 (2002).
\bibitem{hajjoul2013}  H. Hajjoul \textit{et al.}, Genome Res. \textbf{23}, 1829 (2013).
\bibitem{cui2000} Y. Cui and C. Bustamante, Proc. Natl. Acad. Sci. U.S.A. \textbf{97}, 127 (2000).
\end{thebibliography}
\end{document}